\documentclass[epjST]{svjour}
 \usepackage[latin1]{inputenc}
\usepackage[english]{babel}
\usepackage{graphicx}

\usepackage{natbib}
\bibpunct[;]{(}{)}{,}{a}{}{;}
\usepackage{xcolor}
\setcitestyle{notesep={ }}
\newcommand{\be}{\begin{equation}}
\newcommand{\dd}{\displaystyle}
\newcommand{\ee}{\end{equation}}
\newcommand{\bea}{\begin{eqnarray}}
\newcommand{\eea}{\end{eqnarray}}

\def\f{\frac}

\begin{document}

\title{
From Florence to Fermions: a historical reconstruction of the origins of Fermi's statistics one hundred years later}
\author{Roberto Casalbuoni\inst{1}\inst{2}\fnmsep\thanks{\email{casalbuoni@fi.infn.it}}, Daniele Dominici\inst{1} \inst{2}\fnmsep\thanks{\email{dominici@fi.infn.it}} 
\institute{Department of Physics and Astronomy, University of Florence,Via G. Sansone 1, 50019 Sesto Fiorentino (FI), Italy \and INFN, Sezione di Firenze, Via G. Sansone 1, 50019 Sesto Fiorentino (FI), Italy }}

\abstract{Aim of this paper is to retrace the path that led the young Enrico Fermi to write his paper on the statistics of an ideal monatomic gas. This discovery originated in his interest, which he had shown since his formative years, in the absolute entropy constant and in the problems he highlighted in Sommerfeld's quantization in the case of identical particle systems. The fundamental step taken by Fermi in writing his work on statistics was to apply the Exclusion Principle, formulated for electrons in an atom and which could therefore have been a pure effect due to dynamics, to a system of non-interacting particles. } %end of abstract
\maketitle


\section{Introduction}
On 7 February 1926, Antonio Garbasso, Director of the Institute of Physics at the Royal University of Florence and Mayor of the city, presented a paper by 25-year-old Enrico Fermi entitled {\it On the Quantization of the Perfect Monoatomic Gas} \citep{fermistat1} for publication in the Proceedings of the Accademia Nazionale dei Lincei. This work proved to be fundamental in many fields of physics and is the basis for numerous modern technological applications. An expanded version in German was received by  Zeitschrift für Physik on 24 March of the same year \citep{fermistat2}. It was Garbasso himself who had offered the young Enrico a teaching position at the University of Florence two years earlier.
One hundred years after the publication of this work, we retrace the path that led Fermi from his early interest in physics to the publication of these articles.

One of the interesting points in the history of science is the origin of scientific discoveries. It is not always possible to determine what happened at the exact moment the idea was formulated, either because of a lack of direct evidence or because, even if such evidence existed, it is not always reliable. In this regard, and as far as
Fermi's theoretical work is concerned, it is interesting to quote Franco Rasetti's comment in a note
relating to Fermi's work on statistics in the Collected Papers 
 \citep{collected}:
{\it``It is unfortunate that little is known of the circumstances that led Fermi to some of his most significant  contributions in theoretical field, in 
contrast to the adequate information that we possess concerning his experimental work. The reasons are fairly obvious. Even when the experiments were not performed in collaboration with other workers, their progress could be followed almost day by day by his colleagues. As a theoretician, Fermi was entirely self-sufficient, part of his work  was done at home in the early hours of the morning, and occasionally his closest associates had little information on the problem that have occupied his mind until he presented to them, in an informal lecture, the finite product of his meditations.''}

In the case of Fermi's statistics, as we shall see, this discovery
has very specific origins and can be traced back to the particular scientific training
of the young Fermi, who arrived at university with a comprehensive education in
classical physics; this allowed him, during his university years, to
deal with problems at the forefront of physics at the time, namely 
General Relativity, Quantum Mechanics and the problem of determining
the entropy constant.

Rasetti recalls \citep{collected}: {\it``There is conclusive evidence to show that Fermi had been concerned with the problem of
the absolute entropy constant at least since January 1924 when he wrote a paper on the quantization of systems 
 containing identical particles \citep{fermi1924}. He had also been discussing these problems with Rasetti several times in the 
 following year. He told much later to Segrè that the division of phase space into finite cells had occupied him very much and that 
 had not Pauli discovered the Exclusion Principle he might have arrived at it in a roundabout way from the entropy constant.''}
 
 A discussion of the analysis of the succession of ideas and works that led Fermi to the formulation of statistics has been carried out in several works, see e.g. \citep{segre,belloni:1994,cordella2000,parisi,perez:2022}; with regard to their temporal genesis, while Segrè \citep{segre} places it during Fermi's stay in Leiden, Cordella and Sebastiani \citep{cordella2000} anticipate it to the Göttingen period. 
 
 The purpose of this article is to reconstruct the genesis of Fermi's article on statistics,  by placing it in the context of the Florentine physics environment in which Fermi worked. Fermi was called to Florence in 1924 because that was the year the University of Florence was founded and new job opportunities opened up. Garbasso played a fundamental role in Fermi's appointment to Arcetri. Fermi's preparation of the Theoretical Physics course, dedicated in 1925/26 to topics in statistical mechanics,  may have stimulated him to resume his interest in the Sackur-Tetrode formula. Fermi's time in Florence was also important for his collaboration on experimental research with Franco Rasetti.
 
In Section 2, we begin by reviewing Fermi's studies during his high school years,
his preparation for admission to the Scuola Normale Superiore in Pisa, and his
scientific interests at the time. We then briefly retrace his time at the Scuola Normale in Section 3 and, in Section 4, his periods in Göttingen and Leiden, which were important for the work that led him to statistics. Section 5 recalls how Fermi was hired by the newly founded University of Florence and Section 6 analyses the fundamental steps of his 1926 work on the statistics of a monatomic gas.

 \section{The formative years}
 
From an early age, Enrico Fermi showed an interest in building mechanical
and electrical toys while playing with his brother Giulio, who was a year older than him and
died at the age of 15 following a routine surgical operation. Enrico soon came into
contact with mathematics and physics thanks to his encounter with engineer Adolfo Amidei, his father's colleague at the railways in Rome, and the friendship that, after Giulio's death, bound him to Enrico Persico. 

Fermi acquired his first elements of mathematical physics from a book purchased at a stall in Campo dei Fiori during one of his many walks with his friend Persico. The book was Elementorum Physicae Matematicae, written in Latin by Andrea Caraffa, an 1840 treatise on mechanics, optics,
acoustics and astronomy with an introduction to mathematics.

Engineer Amidei, who was passionate about scientific subjects,
particularly mathematics, quickly realised the exceptional talents of the
young Enrico and suggested that he read a series of books, which Fermi began in 1914
and finished in 1918, when he completed his secondary school studies. The books that
Amidei had the young Fermi read over the years were:

1) Geometry: {\it The Geometry of Position} by T. Reye (in 1914);

2) Trigonometry: {\it Treatise on Plane and Spherical Trigonometry} by J. A. Serret (in 1914);

3) Algebra: {\it Course in Algebraic Analysis with an Introduction to Infinitesimal Calculus} by E.
Cesàro (in 1915);

4) Analytic Geometry: {\it Lecture Notes} by E. Bianchi, University of Pisa (in 1915);

5) Infinitesimal and Integral Calculus: {\it The Lectures} of Ulisse Dini in Pisa (in 1916);

6) Rational Mechanics: {\it Traité de Mécanique} by S. D. Poisson (in 1917);

7) Logic and Geometry: {\it Geometric Calculus according to H. Grassmann's Ausdehnungslehre}
preceded by G. Peano's operations of deductive logic (in 1918).

In his usual style, Fermi not only studied these texts, but also did
all the exercises that the books proposed. 

In addition to Amidei, another important figure in the education of the young Enrico was Filippo Eredia, physicist and meteorologist, teacher at the Liceo
Classico Umberto I in Rome (now Pilo Albertelli), where he was professor to both Fermi and Persico, on whom he had a profound influence. Under his guidance, the two boys had calibrated a water thermometer. Since this instrument
could only be used if both the height of the
water column and the temperature were measured accurately, Fermi and Persico calculated a formula for calibration that allowed the pressure to be obtained as a function
of the temperature and the height of the water column
\citep{segre}.

Amidei, deeply impressed by the young Fermi's abilities, 
at the end of high school, advised his family to send him to study at the
Scuola Normale in Pisa. The family was
perplexed, but Amidei eventually managed to
convince them. To prepare for the entrance exam to the Scuola Normale, Enrico
spent the summer studying Chwolson's monumental treatise {\it Traité de Physique},
in French in five volumes totalling 4500 pages.
To this end, Fermi went almost daily to the Library of the Central Institute of Meteorology and Geodynamics, thanks to the permission granted to him by his professor Filippo Eredia \citep{segre}.

In the autumn of 1918, for the entrance exam, which took place in Rome, Fermi wrote an essay on physics
{\it Distinctive characteristics of
sound and their causes} in a superlative manner, with a paper
worthy of a university student in his final year of physics.
The examiners were
astonished and passed him with a mark of 10, declaring: {\it``If the regulations allowed us to do so,
we would have given him honours''}. 
His old professor Filippo Eredia was also a member of the examining board for the entrance examination to the Scuola Normale Superiore in Pisa.

The President of the Board, Professor Pittarelli, who taught
Descriptive Geometry at the University of Rome, was amazed by Fermi's paper and
called him in for an interview, at the end of which he said  he had never met a
student like Fermi and predicted a brilliant future for him \citep{segre}.

 \section{The years at the Scuola Normale}
At the Scuola Normale, Fermi became friends with Nello Carrara and Franco Rasetti. Carrara,
a Florentine, a student at the Scuola Normale and a veteran of the First World War, was a year older
than Enrico and they worked together on their degree thesis. In 1956, Carrara
became a professor in Florence, where he founded the Microwave Centre (among other things, he was also
the inventor of the term "microwave").
Franco Rasetti was born in Pozzuolo Umbro in 1901 and enrolled in engineering
in Pisa; he met Fermi while attending university courses, which at the time were common for
physicists and engineers. Rasetti was a special person with a deep love for
the natural sciences. He was also passionate about the mountains and highly skilled
in manual work. In his third year, he switched to physics because he was impressed
by Fermi's intelligence and knowledge of the subject. After graduating, Rasetti
became assistant to Garbasso\footnote{On Garbasso's forward-looking role in the development of modern physics in Florence, see \citep{bm2006,cdm2021,Casalbuoni:2022rzn}.} in Florence and later joined Fermi at the Institute of Physics in Via Panisperna, Rome.

\begin{figure}[h]
\centering
\includegraphics[width=0.7\textwidth]{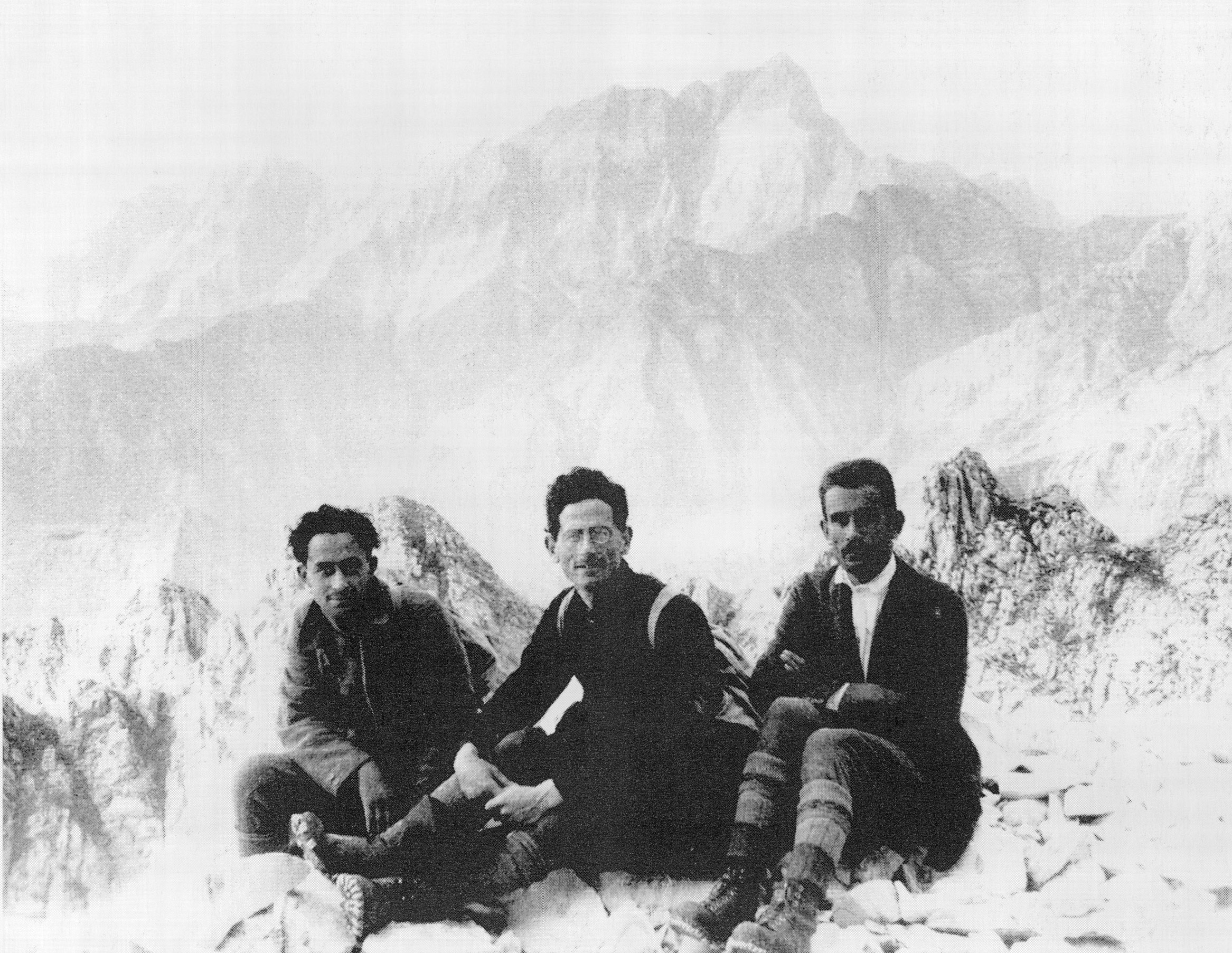}
\caption{The three friends Enrico Fermi, Nello Carrara and Franco Rasetti on a trip to the Apuan Alps.}\label{fig1}
\end{figure}

The three friends were very close, and often, led by Rasetti, they went on
excursions to the Apuan Alps (see Fig. \ref{fig1}).

Fermi's background in physics and mathematics meant that he did not need to study the
topics covered in the degree course and was able to help his friends Carrara and
Rasetti pass their exams, and to study cutting-edge topics such as General Relativity
and the quantum theories of Bohr and Sommerfeld. His education was enriched by A. Sommerfeld's book, 
{\it  Atombau und Spektrallinien}, and { \it Electron Theory of Matter} by O.W. Richardson.
The knowledge of most Italian physicists at the time was limited to classical physics,
and Fermi was practically the only Italian physicist to know
these two disciplines, apart from a few mathematical physicists such as Levi Civita. 

In particular, General Relativity was only followed by mathematicians, given
its deep connection with geometry. It was precisely in this field
that Fermi, in 1922, before graduating, published an important work \citep{fermi1922} in which
he introduced a particular coordinate system that was also taken up by Walker \citep{walker1932,walker1932a}, 
coordinates now known as Fermi-Walker coordinates.

Fermi's abilities are attested to by the words of Carrara \citep{carrara}:
{\it``Fermi was not a student like the others: he did not study any lessons, and he did not
need to; he went his own way. While we were forced to struggle to
keep up with the teachers, he was already dealing with the most current problems in physics.
Who among us knew anything about that mysterious character in thermodynamics known
as entropy? No one; but Fermi already had something new to say
about it.''}

Fermi's teachers soon realised the student's abilities and, one day,
the Director of the Institute of Physics in Pisa, Luigi Puccianti, asked him to teach him
something about quantum theories, saying: {\it``Because I am a dunce, but if
you explain things to me, I understand them''} \citep{maltese}. In fact, Fermi had the ability to explain
the most complex things in simple terms.

Fermi graduated on 4 July 1922 with an experimental thesis on the formation of images with 
X-rays and, a few days later, on the 7th, he discussed his habilitation thesis with a
paper entitled: { \it A theorem of probability calculation and some applications}.

\section{The first attempts: from Göttingen to Leiden, 1923-1924}

After graduating, Fermi returned to Rome where he introduced himself to the Director of the Institute
of Physics at the Royal University of Rome, Senator Orso Mario Corbino, a very influential person
in Italy, both scientifically and politically. Corbino, who had long been thinking about
how to revive Italian physics, immediately realised that Fermi
was absolutely extraordinary and saw in him the possibility of creating a
school of physics that could put Italy on a par with other European nations. He therefore
encouraged the young Fermi to go to Göttingen, which at the time was perhaps the most
important centre in the world for theoretical physics, with a scholarship for study abroad from the Ministry of Education. At the Göttingen Institute there were people of the calibre of Max Born and
Werner Heisenberg (both future Nobel Prize winners) and Pascual Jordan (a central figure
in the development of quantum mechanics).
Almost all of Fermi's biographers argue that this period was not a happy one
for the young physicist. While he was held in very high regard in Pisa, in
Göttingen there were other young people and personalities who were just as brilliant. Considering also his
youthful shyness, all these factors probably contributed to his isolation.  Particularly interesting to this theme is the oral history interview to Heisenberg by Kuhn \citep{kuhn-heisenberg}. Heisenberg states:{\it``Born appeared to me, when I came to Göttingen, as an extremely good
mathematician who was interested in the mathematical methods of physics, but who did
not know so many details. He had not so much feeling about how the things in atomic
physics were. So I remember that, for instance, Fermi, who took part in this seminar was
a bit unhappy about it; he disliked these mathematical subtleties, proof of convergence,
and such. For example, we spoke about the theorem of (Bruins) that in an infinite
neighborhood of a periodic solution there are always solutions which are not periodic;
this kind of mathematical subtlety he hated. I mean, Fermi felt, ``That's not physics''. ''} According to Heisenberg, Fermi could speak 
 some German, but in some way he was a bit shy, and perhaps he didn't like
the whole atmosphere of the country.

Nevertheless, during this period he wrote several papers on analytical mechanics that
earned him the interest of Paul Ehrenfest\footnote{Ehrenfest had noticed Fermi's work on the ergodic theorem \citep{fermi1923nc,fermi1923zf,fermi1924zf} and had invited his student George Uhlenbeck, who was in Rome, to meet Fermi.} and became aware of issues that led him to write articles on statistics.

  For example, he learned
of the enormous difficulties encountered by Born in applying 
Sommerfeld quantization to the helium atom and probably of a paper by Stern \citep{stern} on the
derivation of the formula by Sackur \citep{sackur,sackur2} and Tetrode \citep{tetrode1912}\footnote{See, in this regard, \citep{cordella2000}.}.

The work of Sackur and Tetrode dealt with the
problem of the absolute constant of entropy of a perfect gas which, in classical statistical mechanics, is defined to within a constant:
\begin{equation}
S=kN \left (\frac 3 2 \log \frac E N +\log  \frac V N +s_0\right ),
\end{equation}
where $E, V $ and $ N$ are respectively the energy, volume and number of atoms, and the constant $s_0$ is an indeterminate constant.

Sackur and Tetrode, independently\footnote{For a detailed historical review, see \citep{grimus}.},
had derived this constant using a Boltzmann-like description, calculating entropy as

\begin{equation}
S=k\log W,
\end{equation}
where $W$ is the number of possibilities for realising the macroscopic state
but assuming
a hypothesis of quantization of phase space, i.e. that it was divided
into cells of volume proportional to Planck's constant cubed
\begin{equation}
\delta q_1\delta p_1...\delta q_n\delta p_n\sim (zh)^n,
\end{equation}
with $n=3N$ and $z$ a dimensionless constant.
The value of $s_0$ was
\begin{equation}
s_0=\frac 3 2 \log\frac {4\pi m}{3 (zh)^2}+ \frac 5 2,
\end{equation}
where $m$ is the mass of the atom.

Both were also able
to demonstrate that the formula was in agreement with experimental data relating to the vapour pressure 
of mercury if $z\sim 1$. The nucleus of the most common stable state of mercury is composed of 80 protons and 121 neutrons and is therefore a fermion.

The quantization hypothesis of these authors was arbitrary and therefore
numerous researchers devoted themselves to alternative derivations of this formula, including
Stern in his aforementioned work. It should be emphasised that, according to
Nello Carrara, in a lecture given in 1955 \citep{carrara}, Fermi, since
he was a student at the Normale, had been interested in the problem of the entropy constant.
In the words of Carrara  \citep{carrara}:{\it``. . . the determination of the absolute value of entropy
was the problem that swirled in the mind of the young Fermi, not yet twenty years old,
who ate castagnaccio with us, climbed the statue of Cosimo I, and made sacrifices
to entropy in the Apuan Alps; . . . ''}

In any case, Fermi's dissatisfaction with his stay in Germany meant that, instead of
staying for a year, he ended his visit in August 1923 and returned to Rome. Here,
Corbino obtained him a teaching position in Mathematics for Chemists for the academic year
1923/24. During this period in Rome, he wrote two papers on statistical mechanics that
were the precursors to the formulation of statistical mechanics in 1926. The first, entitled
{\it  On Stern's theory of the absolute constant of the entropy of a perfect
monoatomic gas} \citep{fermi1923}, presented to the Accademia dei Lincei by Corbino on 2 December
1923, the second: {\it  Considerations on the quantization of systems containing
identical elements } \citep{fermi1924}, sent to Il Nuovo Cimento in January 1924. 

The first of these two works represented an improvement on Stern's formulation, which considered a gas as a vapour of a solid body and calculated its maximum vapour pressure using two methods: kinetic theory of gases and thermodynamics. From the comparison, he determined the absolute entropy constant, but assuming, in the thermodynamic calculation of the saturated vapour density, an energy at absolute zero for each oscillator equal to $h\nu/2$. Fermi demonstrated in his work that this assumption was not necessary, but in the kinetic calculation of saturated vapour density, he assumed that the energies of the molecules of the solid were equal to 
$n_j h\nu_j,\quad j=1... 3N_s$ where $N_s$ 
is the number of molecules in the solid, with $n_j$ an integer and $\nu_j$ the characteristic frequency of the molecule, and that $dq_jdp_j=h$ for each molecule in the solid. In this way, the kinetic calculation of the saturated vapour density led to the same result as the thermodynamic calculation.

\begin{figure}[h]
\centering
\includegraphics[width=0.7\textwidth]{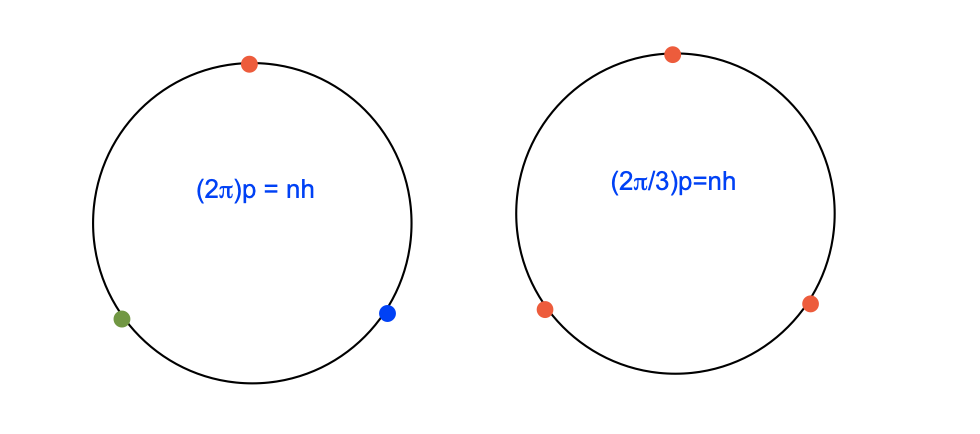}
\caption{Illustration of Sommerfeld's quantization rules for three distinguishable particles (on the
left) and three identical particles (on the right).}
\label{fig2}
\end{figure}

The
second is more interesting for reconstructing the path that led the young Roman scientist to his work on the quantization of monatomic gas. First, Fermi
showed how Sommerfeld's quantization rules 
\be
\oint p_i dq_i=n h
\ee
gave rise to different results
in systems with distinguishable parts and in those with identical parts. Fermi
illustrated this point by considering a ring of unit radius occupied by three particles placed at the vertices of an equilateral triangle, as in Fig.\ref{fig2} (Figures 2 to 4 are not present in the paper \citep{fermi1924} but have been inserted to clarify Fermi's reasoning). The
system has different periodicities; in the case of three distinguishable particles, the periodicity is equal to 2 $\pi$, while in the case of three identical particles, it is equal to
2/3 $\pi$. Therefore, if $p$ is the momentum of the system, by using the Sommerfeld's rules in the first case $2\pi p$ is a multiple of $h$, while in the second case $2\pi /3\, p$ is a multiple of $h$,
which therefore results in a different quantization of momentum.

\begin{figure}[h]
\centering
\includegraphics[width=0.4\textwidth]{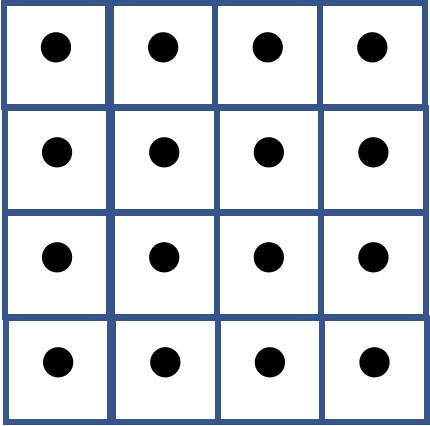}
\includegraphics[width=0.4\textwidth]{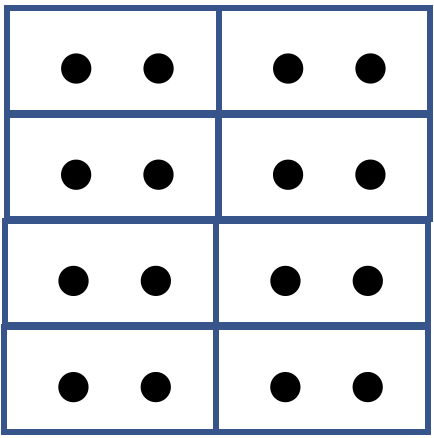}
\caption{Two-dimensional representation of the volume of gas divided into $N$ cells with one molecule per cell (left) and into $N/2$ cells with two molecules per cell (right).}\label{fig3}
\end{figure}

In the central part of this work, Fermi took up an interesting contribution
by Brody \citep{brody} for calculating the entropy of a perfect gas. Fermi considered a gas composed of N
particles placed in a parallelepiped-shaped container, then divided the container
into N cells and placed one particle in each cell, as in Fig.~\ref{fig3} on the left. The momentum of each particle was
quantized according to Sommerfeld's rules. Each cell was characterised by the
three quantum numbers that define momentum in the three spatial dimensions. In this
way, the author was able to reproduce the formula of Sackur-Tetrode. Fermi also considered
the possibility of dividing the parallelepiped into N/2 cells, each containing 2
particles (Fig.~\ref{fig3} on the right) and so on until it was reduced to the parallelepiped itself with N particles,
always using Sommerfeld quantization. In all these cases, the final formula
for entropy differed from that of Sackur-Tetrode.

\begin{figure}[h]
\centering
\includegraphics[width=0.4\textwidth]{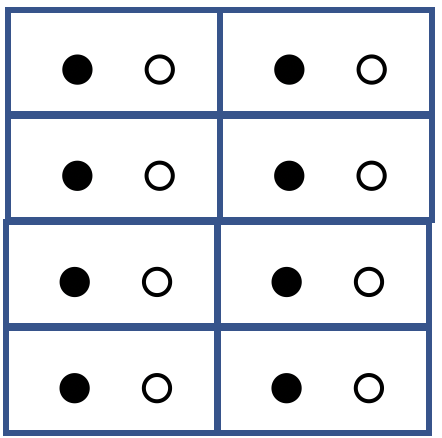}
\caption{Two-dimensional representation of the volume of gas divided into $N/2$ cells with two different molecules per cell. }\label{fig4}
\end{figure}

Fermi concluded his work
by attributing the reason for the discrepancy with Sackur and Tetrode, in the case of multiple particles
in the same cell, to the failure of Sommerfeld's quantization rules in the case of identical particles. To better reinforce this statement, Fermi considered a mixture of two gases, i.e., a set
of N particles divided into two groups of N/2 distinct particles but with equal mass. He then divided the parallelepiped into N/2 cells and placed
one particle of one type and one of the other type in each of them (Fig.~\ref{fig4}). He then showed that the result
was as predicted by thermodynamics, i.e. that the entropy was equal to the sum
of the entropies of the two gases, as if each occupied the total volume, with the individual
entropies coinciding with those found by Sackur and Tetrode.

 Fermi made no further comments and did not even offer a possible explanation, which
he would later find to be the Exclusion Principle. However, in hindsight, Fermi complained
to Segrè \citep{segre} that he had not been able, on the basis of these arguments, to formulate the Exclusion Principle before
Pauli.

According to Bruno Pontecorvo, {\it``Fermi had been toying with the idea of this work for some time, but he lacked Pauli's principle. As soon as the latter was formulated, he sent his article to press. In this regard, it must be said that Fermi was rather bitter about not having been able to formulate Pauli's principle on his own, a principle to which, as his work shows, he had come very close''} \citep{pontecorvo1993}.

In the autumn of 1924, Fermi, with a grant from the Rockefeller Foundation, 
went to Leiden to join Ehrenfest, an environment more congenial to him both in terms of the scientific interests of the group and the atmosphere. Among Ehrenfest's students were Uhlenbeck and Samuel Goudsmit, who together proposed the electron spin \citep{Uhlenbeck:1925pqz}. Furthermore, Fermi had the opportunity during that period to meet Lorentz and Einstein, who were visiting Leiden.

 \section{The Florentine period}

\begin{figure}[h]
\centering
\includegraphics[width=0.45\textwidth]{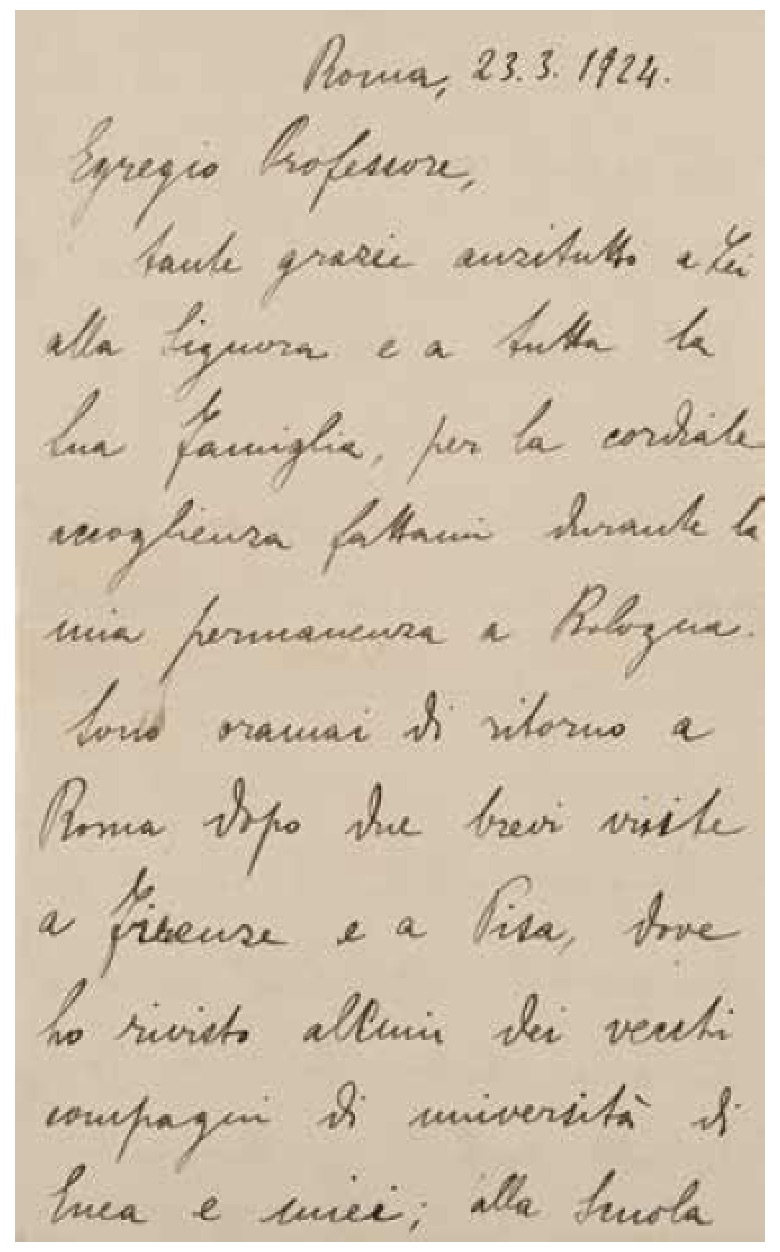}
\includegraphics[width=0.465\textwidth]{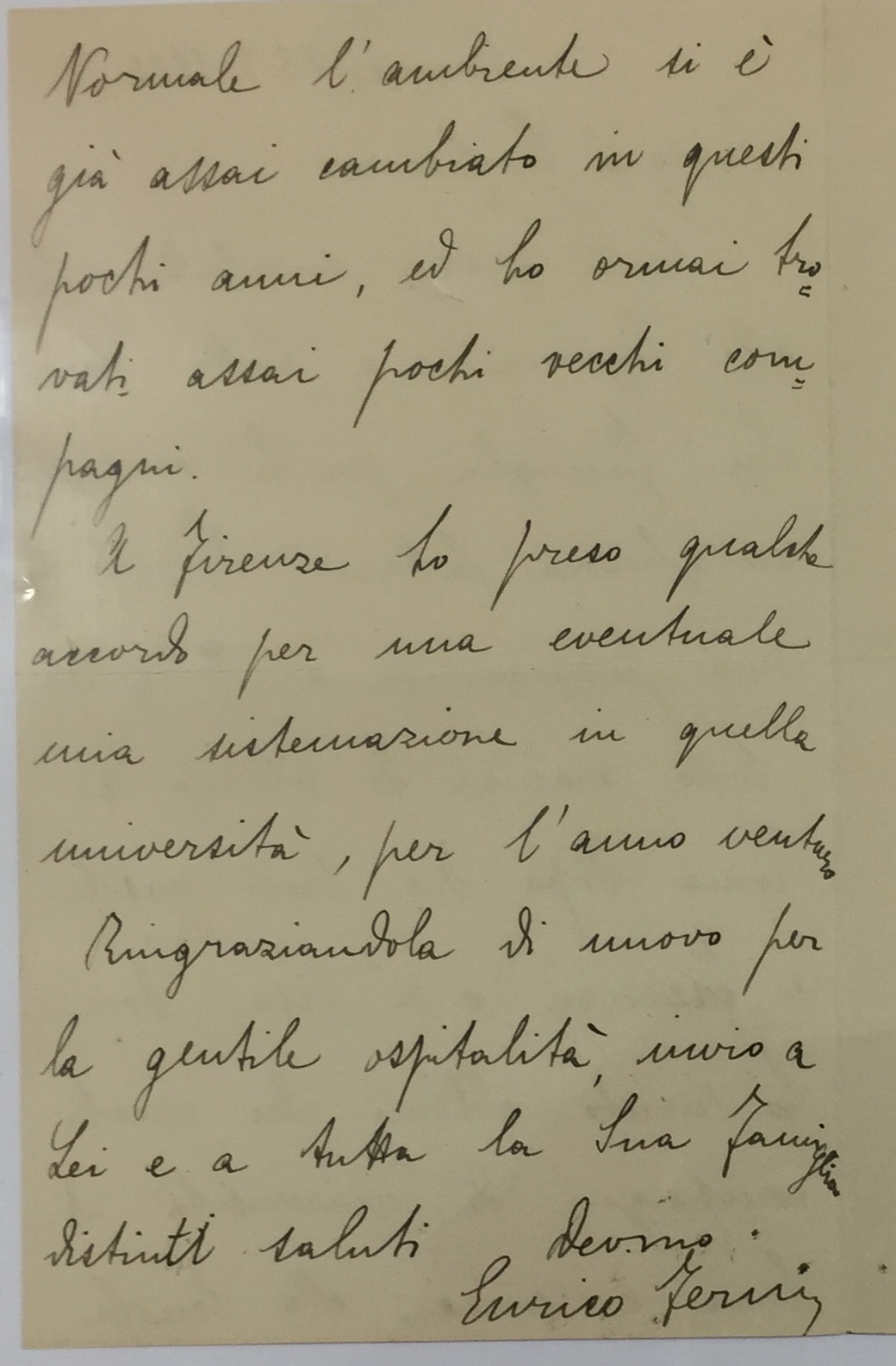}
\caption{Fermi's letter to Ettore Bortolotti, from  \citep{guerrarobotti}.}\label{fig5}
\end{figure}

Meanwhile, in a letter dated 23 March 1924 \citep{guerrarobotti} addressed to Ettore Bortolotti, Professor
of Geometry at the Royal University of Bologna and father of Enea, Enrico's classmate in Pisa, Fermi wrote
{\it``In Florence, I have made some arrangements for my possible placement at
that university for the coming year''} (Fig.~\ref{fig5}). Garbasso was therefore the first to offer Fermi a temporary physics position at the newly established Royal University of Florence in the academic year 1924/25, which was renewed the following year. 

Antonio Garbasso had been appointed to the chair of Experimental Physics at the Institute of Higher Practical Studies and Specialisation in Florence in 1913. He graduated in Turin in 1892 with Andrea Naccari and in 1893-94 he went to Germany to continue his experimental studies on optics and the electromagnetic theory of light, first for a semester at the Institute of Physics in Bonn with Heinrich Hertz and then for another six months in Berlin, collaborating with Hermann von Helmholtz on radiative absorption and resonance processes. 
After teaching courses in Turin and Pisa, in 1903 he was awarded the chair of Experimental Physics and was called to Genoa. In Florence, Garbasso promoted the development of physics by building the new laboratory headquarters in Arcetri and founding a school open to the new trends in physics at the beginning of the century, calling on promising young physicists such as Rasetti and Fermi. Among Garbasso's achievements 
is that of being the first Italian physicist to use Bohr's quantum theory. Garbasso used it in his explanation of the Stark-Lo Surdo effect, a work presented to the Accademia dei Lincei in 1913 \citep{leone}.

During the years when Fermi was teaching in Florence, Garbasso's assistants included, in addition to Rasetti, Rita Brunetti\footnote{Rita Brunetti (1890-1942). She graduated in Physics from Pisa in 1913, with Augusto Occhialini as her supervisor, and was the first woman to win a chair in Physics in Italy, in 1926.}, who worked in spectroscopy in the visible and X regions, and Vasco Ronchi\footnote{Vasco Ronchi (1897-1988). Graduated in Physics at Pisa in 1919, supervisor Garbasso. From 1927 Director of the Laboratory of Optics and Mechanics in Florence, which in 1930 became the National Institute of Optics.}, expert in optics. Antonino Lo Surdo had been an assistant at Arcetri from 1908 to 1916, and it was here that, at Garbasso's suggestion, he carried out the experiment that led to the discovery of the Stark-Lo Surdo effect. Augusto Occhialini had been Garbasso's assistant from 1918 to 1921. After conducting several experiments on the Stark-Lo Surdo effect in helium, Rita Brunetti carried out a number of experiments on the behaviour of high-frequency spectra in magnetic fields in 1918-20. In two papers on X-ray production, she used Bohr's quantization hypothesis to interpret the results obtained. In 1921, she wrote a review paper on Bohr's quantum model.

The Mathematical Physics course, which Fermi taught in the fourth year of the degree courses in Physics, Physics and Mathematics, and Mathematics, covered the traditional topics of Electrodynamics for the 1924/25 academic year, supplemented by a mention of the new Theory of Relativity.
The following year, the title of the course was changed to Theoretical Physics, and Fermi covered concepts of probability, thermodynamics, and statistical mechanics.
The second course he taught, Rational Mechanics, was aimed at physicists and engineers. The course notes, collected by two engineering students, have recently been republished \citep{cdp2019}.

\begin{figure}[h]\centering
\includegraphics[width=0.6\textwidth]{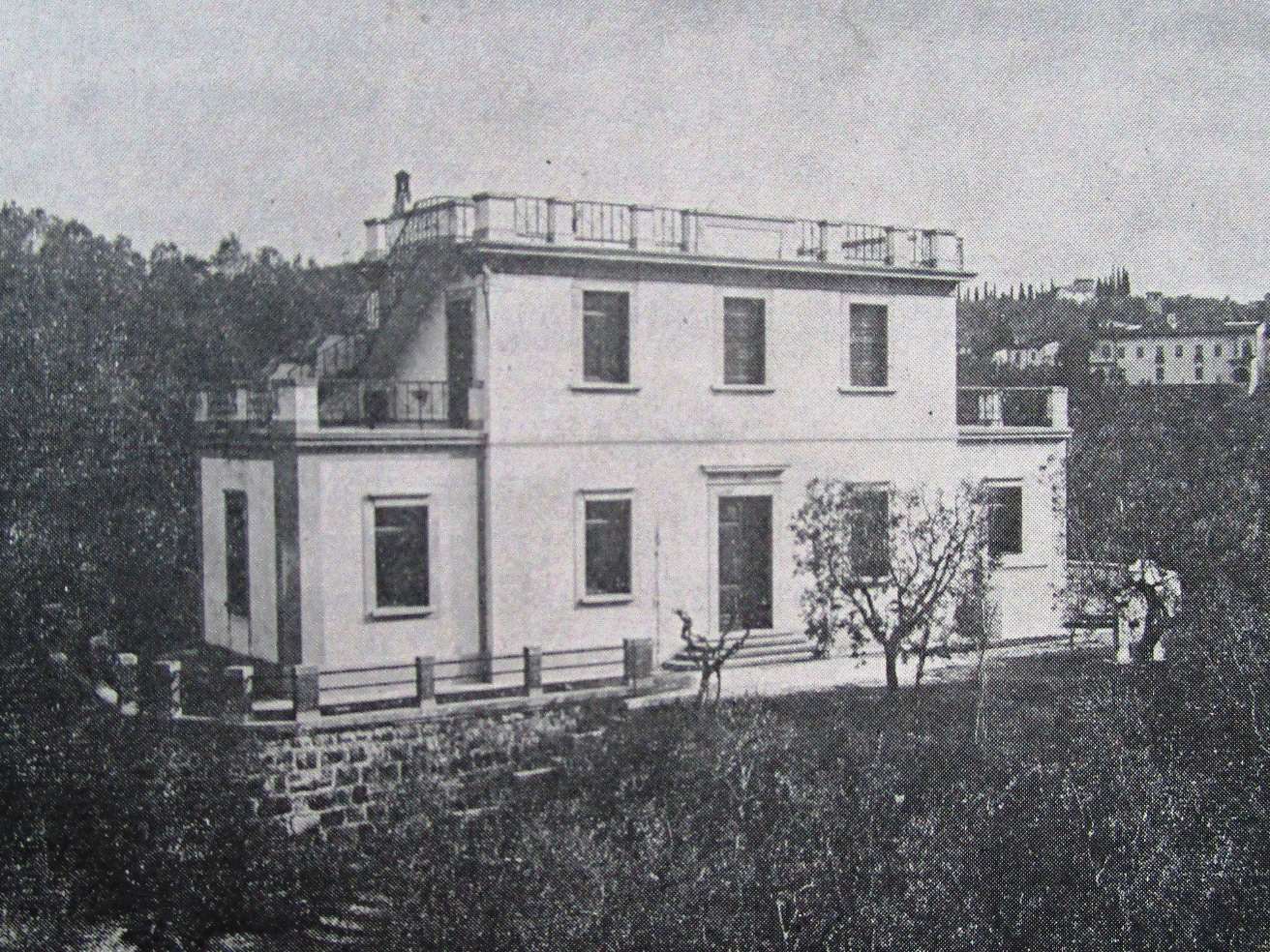}
\caption{The {\it vagoncino}, originally the headquarters of the Institute of Earth Physics, was used as accommodation first by Rasetti and then by Fermi; it was also the headquarters of the Laboratory of Optics and Precision Mechanics (from the Proceedings of the Giorgio Ronchi Foundation).}\label{fig6}
\end{figure}

During his time in Florence, Fermi stayed in the so-called {\it vagoncino} (Fig.~\ref{fig6}), the premises that later became the first headquarters of the National Institute of Optics in Arcetri. In that building, there was a room with a bed and a stove that had allowed Rasetti, in the previous two years, to find accommodation in the company of scorpions \citep{goods}. After the death of Rasetti's father and his mother's move from Pisa to Florence, Rasetti moved in with her, leaving the room in the {\it vagoncino} to his friend.

In Florence, Fermi also began to worry about his academic future, as evidenced by his correspondence with Persico. In 1925, he obtained the  
{\it Libera docenza} in Mathematical Physics and attempted to compete for his first professorship in Mathematical Physics at the University of Cagliari. This chair was awarded by majority vote to Giovanni Giorgi, supported by three members of the Committee, Giovanni Guglielmo, Roberto Marcolongo and Carlo Somigliana; Fermi was supported by the mathematical physicists Tullio Levi Civita and Vito Volterra, who were more aware than the other members of the importance of the new physics. In particular they had a good appreciation of Fermi also for his work on General Relativity.

Furthermore, in view of his possible appointment to the first Italian chair of Theoretical Physics in Rome, which took place in the autumn of 1926, Fermi took steps to be replaced by Persico in the teaching of Theoretical Physics in Florence. In fact, he wrote to his friend in a letter dated 13 May 1925
{\it``Dear Enrico, thank you for your postcard. I have learned that a competition for Theoretical Physics has been announced in Rome, so the possibility of having to think about my succession here is becoming a reality. All of us who would like you to be my successor, namely Prof. Garbasso, Tricomi and myself, do not know whether you would prefer the competition to be announced or to be appointed as my successor in due course...''}. The competition in Rome, delayed by a year for reasons related to possible changes in the rules governing university competitions, was concluded in the autumn of 1926; Fermi was appointed to Rome, Persico (second in the competition) was appointed to Florence on 1 January 1927 \footnote{Aldo Pontremoli (third in the competition) was appointed  to Milan. Unfortunately Pontremoli died in 1928, after  the tragic crash of airship Italia during the mission of Nobile to the North Pole \citep{colombi:2021}.}.

 With the arrival of Fermi to Rome and Persico to Firenze a strong  boost was given   to  the project of Corbino and Garbasso of creating an Italian school of modern physics  from both the experimental and theoretical point of view.

\section{Fermi's statistics}

Upon his arrival in Florence in December 1924, Fermi reunited with his friend Franco Rasetti, a highly respected experimental physicist, who involved him in some experimental research, the first since his thesis on X-rays.
This research kept him busy until the spring of
1925, analysing the effect of weak but high-frequency magnetic fields on the depolarisation of resonance light in mercury vapour \citep{rasfermi1925,rasfermi1925a}. This work is the first example of the study of atomic spectra using radiofrequency fields, a technique that would receive numerous applications in subsequent years.

Meanwhile, in January of the same year, Pauli's work was published \citep{pauli}
with the formulation of the famous Exclusion Principle, which excluded the coexistence
of two electrons in the same atomic state. It was not at all clear whether this
was a principle of nature or a consequence of atomic dynamics.
It is curious that it took Fermi a year, from the formulation of Pauli's Principle
to his work in 1926 \citep{fermistat1,fermistat2}. An explanation has been proposed by Cordella and Sebastiani \citep{cordella2000},
who observed, as we have already noted, that Fermi had been busy until the spring of 1925 with
experimental work with Rasetti. In July and September, he was busy as a
member of the Committe  for the state examinations in Florence. In August, as he did systematically
every year, he went on holiday to San Vito di Cadore. Here he met some friends,
including R. de L. Kronig. Kronig was a theoretical physicist who was very well informed about recent developments
in physics and was particularly aware of Pauli's article, as
can be seen from a letter he wrote to Goudsmit in January 1925. It is therefore plausible
that Kronig reported this to Fermi. Evidently, the Roman physicist was not
very impressed by this news, given that on 23 September 1925 he wrote to his friend
Persico saying that he had only recently resumed reading some scientific works,
but his feeling was that there had been no particularly interesting results.
According to Cordella and Sebastiani \citep{cordella2000}, there are no other reliable sources until February 1926,
except for a letter to Kronig dated 23 November 1925 in which he wrote:
{\it``I have now almost finished the theory of heat conduction in crystals
(you may remember that I was already working on it in S. Vito)... I would like to ask you if you are
aware of any publications by Pauli on this subject, as I have not been able to find anything in the literature
... I have started my lectures here. This year I am teaching probability calculus
and statistical mechanics.''} Fermi's reference to the
course in statistical mechanics he was preparing suggests that he may have regained
interest in the subject of the entropy of a perfect gas on that occasion.

We can also add that Laura Fermi, in her book {\it Atoms in the Family} \citep{caponl},
writes that Fermi and Rasetti used to go hunting for lizards in the meadows of Arcetri and that
Fermi, while lying in the grass, waiting to catch them with a glass rod
with a loop at the end, {\it``... let his mind wander. His subconscious
worked on the Pauli Principle and the theory of perfect gas. From the depths of his
subconscious came the missing factor that Fermi had long sought...''}

According to Giulio Maltese \citep{maltese}, in a radio interview given the day after
the Nobel Prize was awarded, when asked if his 1926 work was due to a sudden flash of inspiration,
Fermi replied: {\it``... I remember, for example, that the first idea on how to treat
the statistics of electron gases came to me while I was walking in Arcetri on a
December afternoon next to the Astronomical Observatory.''}

Ultimately, giving credence to Laura Fermi's testimony and Giulio Maltese's account,
after Fermi's subconscious had been working throughout the autumn,
the missing factor finally emerged clearly in December 1925.

It should be remembered that Fermi's work was carried out within the framework of the
old quantum theory, so the only theoretical means of quantizing a system
was to resort to Sommerfeld's rules. It should be noted that Heisenberg's matrix mechanics
had been formulated in July 1925 \citep{Heisenberg:1925zz}. Fermi was familiar with
Heisenberg's work, but considered it rather formal and unsuitable for
use in meaningful applications. Wave mechanics, on the other hand, was formulated by
Schrödinger in March 1926 \citep{Schrodinger:1926xyk}; so Fermi's work could only consist of
applying the statistical methods implemented with the Sommerfeld quantization method.

Fermi began his 1926 article \citep{fermistat1} by recalling the problems associated with
Sommerfeld quantization in the case of identical particles. He then went on to consider
the Exclusion Principle and generalised it by requiring that it apply not
only to identical particles in interaction, such as electrons in an atom, but
also to non-interacting particles, such as atoms in a perfect gas. In other words,
Fermi assumed that the Exclusion Principle was, for the particles that satisfy it,
an intrinsic characteristic and not due to external factors, such as the interactions
to which they might be subjected. At this point, he could calculate
the entropy of the perfect gas as he had done in his 1924 work \citep{fermi1924}, but to be more
consistent with the Exclusion Principle as formulated by Pauli, he required that each molecule, at a distance $r$ from a centre O, be acted upon 
a harmonic potential so that the energy $w $ of the perfect gas was quantized and
characterised by the 3 integers $ s_i, i = 1, 2, 3$ according to the relation:
\begin{equation}
 w = (s_1 + s_2 + s_3)h\nu\equiv  sh\nu
 \end{equation}

Therefore, according to Pauli's Principle, 
the value $sh\nu $ can be taken at most from
\be
 Q_s=\f 1 2 (s+1)(s+2)
 \ee
molecules.
Assuming we have $N$ molecules at absolute zero temperature, he observed that if $N=1$, the only molecule will have $s=0$, if $N=4$ one molecule will occupy the level with $s=0$ and three the levels with $s=1$ and so on. He then indicated with $N_s\leq Q_s$ the number of molecules with energy $s h\nu $, and with $W=Eh\nu$ the total energy, then
 \be
  N=\sum_s N_s,\quad E= \sum_s s N_s
  \label{condizioni}
  \ee
and calculated the probability 
of realisation $P$ of a state with $N_0$ molecules in state $s=0$, $N_1$ molecules in state $s=1$...up to $N_s$ molecules in state $s$. The details of the calculation of $P$ are given in the second paper \citep{fermistat2}:
\be
  P=\prod_s \pmatrix{Q_s\cr N_s}=\prod_s \f {Q_s!}{(Q_s-N_s)!N_s!}
  \ee
We have
  \be
  \log P=\log \prod_s \pmatrix{Q_s\cr N_s}=-\sum_s\left ( N_s\log \f {N_s}{Q_s-N_s}+Q_s\log \f {Q_s-N_s}{Q_s}\right )
  \ee
  where in the last step Stirling's formula $n!\sim n^n e^{-n}$ was used. He then calculated the most probable distribution, maximising $\log P$, taking into account the two conditions (\ref{condizioni}) and finding
  \be
  N_s=Q_s \f {\alpha e^{-\beta s}}{1+\alpha e^{-\beta s}}
  \label{distrib}
  \ee
  with $\alpha$ and $\beta$ constants dependent on $W$, $T$ and $N$.
To calculate $\beta$, he observed that for $r\to\infty$ the density of molecules must tend to zero, so any degeneration effect disappears and the distribution (\ref{distrib}) must tend towards the Maxwell distribution. This requirement leads to the determination of $\beta$, demonstrated in the second paper \citep{fermistat2}, 
 \be
  \beta=\f {h\nu}{kT}
  \ee
Fermi then derived the gas pressure, the specific heat at low temperatures with the correct linear behaviour in $T$ for $T\to 0$ and the expression of entropy at high temperature, rediscovering the formula of Sackur and Tetrode.

It is worth noting that Fermi applies the pressure formula to a gas with the atomic weight of helium, which is a boson and does not satisfy the Exclusion Principle, because, as we will see later, Dirac will also do something similar in his work on statistics.

In his derivation of statistics, Fermi does not seem to have been influenced by the work on photon statistics by Bose  \citep{Bose:1924mk}, received by the journal on 2 July 1924, and that of Einstein  \citep{einstein:1924} of 10 July 1924, although the latter is cited. Einstein and Bose, in their derivation of statistics, require the quantization of cells in phase space, whereas Fermi considers quantized energies by treating molecules as quantum harmonic oscillators. The introduction of absolute temperature is also different  \citep{perez:2022}.  

We agree with Pérez and Ibáñez's conclusion  that Fermi's work cannot be considered foundational, being formulated in the framework of the old quantum mechanics ideas. However its contribution to the physics is enormous. We also note that Fermi knew very well the statistical approach, as shown in  \citep{fermi1924}, where he recovered the Sackur-Tetrode formula, assuming  one particle per cell. In this sense Fermi was very close to the Exclusion Principle.

\begin{figure}[h]\centering
\includegraphics[width=0.65\textwidth]{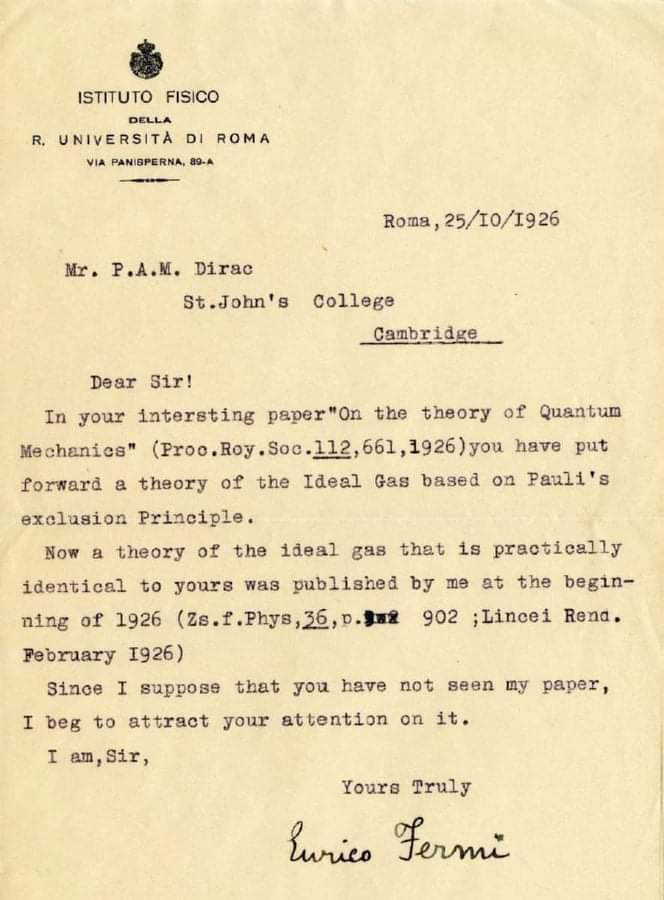}
\caption{The letter that  Fermi wrote to  Dirac (Credit: Luisa Cifarelli).}
\label{fig7}
\end{figure}

  The work on monatomic gas was the last important work of the old quantum theory period.
Shortly afterwards, in August 1926, Dirac \citep{Dirac:1926jz} showed how, starting from
wave mechanics, identical particles could be divided into two broad categories,
respectively with symmetric or antisymmetric wave functions with respect to the exchange of the coordinates of two identical particles, that is, in the modern terminology subsequently proposed by Dirac himself, into bosons and
fermions, and he derived the corresponding statistical distributions. Since Dirac did not cite the work of the Roman scientist in his article, Fermi wrote him a letter, Fig.~\ref{fig7}, perhaps at Corbino's suggestion \citep{belloni:1994,segreejp}, pointing out the priority of his own work.
Very honestly, Dirac recognised the priority of Fermi's work, and so the statistics
of particles that satisfy the Exclusion Principle took the name of Fermi-Dirac statistics. Furthermore, in a talk given in 1945 at Palais de la Découverte in Paris, Dirac, in honour of Fermi, introduced the word ``fermion'' to define the particles satisfying the Fermi-Dirac statistics and ``boson'' to define those satisfying Bose-Einstein statistics \citep{Farmelo2009TheStrangestMan}.

Fermi had written his work in his usual style, introducing general principles
but applying them to specific cases. In fact, Fermi did not like a theory as such,
but only appreciated it if it was able to explain experimental facts. Dirac, on the
contrary, loved the formulation of elegant theories and, even if he read Fermi's work,
it is likely that he considered it a paper devoted to applications. In an interview with Kuhn \citep{kuhn2}, Dirac stated that he had read the work but then forgot about it. It is curious that in his work \citep{Dirac:1926jz}, after deriving the formula,
\be
N_s=\f {A_s}{e^{ \alpha+E_s/kT}+1}
\ee
 Dirac writes {\it``This formula gives the distribution function of the molecules''}, exactly as Fermi applies his statistics to the helium atom. In his work we find again {\it``The solution with symmetrical eigenfunction must be the correct one when applied to light quanta, since it is known that the Einstein-Bose statistical mechanics leads to Planck's law of black-body radiation. The solution with anti-symmetrical eigenfunctions, though, is probably the correct one for gas molecules since it is known to be the correct one for electrons in an atom and one would expect molecules to resemble electrons more closely than light quanta''.}
To contextualise this statement by Dirac and Fermi's application to $^ 4$He gas, which is a boson, it is worth remembering that at the time the only known particles were the proton and the electron, constituents of matter, and the photon, the quantum of electromagnetic waves.
As Kragh also points out \citep{kragh}, { \it``...the belief that gas molecules satisfy the same statistics as electrons was not peculiar to Dirac; it was rather generally assumed in 1926 and was, for example, also a part of Fermi's early work on quantum statistics.''}

\section{Closing remarks}

Fermi-Dirac statistics was immediately appreciated and applied, by
Fowler to stellar matter \citep{fowler1926} and  by Pauli to explain the weak paramagnetism of alkali metals \citep{pauli1927}\footnote{A detailed discussion of  early applications of Fermi-Dirac  statistics is presented in \citep{perez:2022}}.
 An extended discussion from  these first applications of a degenerate gas of  electrons to the more general context of dense matter in the stars, including neutron stars can be found in  \citep{Bonolis:2017fdf}.

In September 1927,  on the occasion of the 100th anniversary of Alessandro Volta's death, an international conference was organised in Como that  turned out to be very  relevant for the progress of Quantum Mechanics and quantum statistics. Several Nobel prizes were present, F. W.  Aston, N. Bohr, P. Bragg, A. H. Compton, J. Franck,  M. v. Laue, H. A. Lorentz, R.A. Millikan, M. Planck, E. Rutherford, P.  Zeeman; also present were M. Born, L. De Broglie,  J. Frenkel, W. Heisenberg,  W. Pauli, A. Sommerfeld and O. Stern. Italy was represented by  Fermi, Corbino, A. Garbasso, Levi-Civita, Volterra and others. Como is the place where
 Bohr discussed for the first time his preliminary  ideas on complementarity, that he presented also, in an improved version, one month later at the Fifth Solvay Conference in Brussels. 
Sommerfeld emphasised the importance of Fermi's work in explaining the properties of metals that had been difficult to understand until then. The title of his talk {\it``Zur Elektronentheorie der Metalle und des Volta-Effektes nach Fermi-schen Statistik''} explicitly quotes ``Fermi statistics''. Results of his work were published as \citep{sommerfeld1927}. 
During the discussion after the Bohr talk, Fermi outlined the important physical applications of his quantum statistics, by illustrating  the results of Pauli and Sommerfeld.

\begin{figure}[h]\centering
\includegraphics[width=0.4\textwidth]{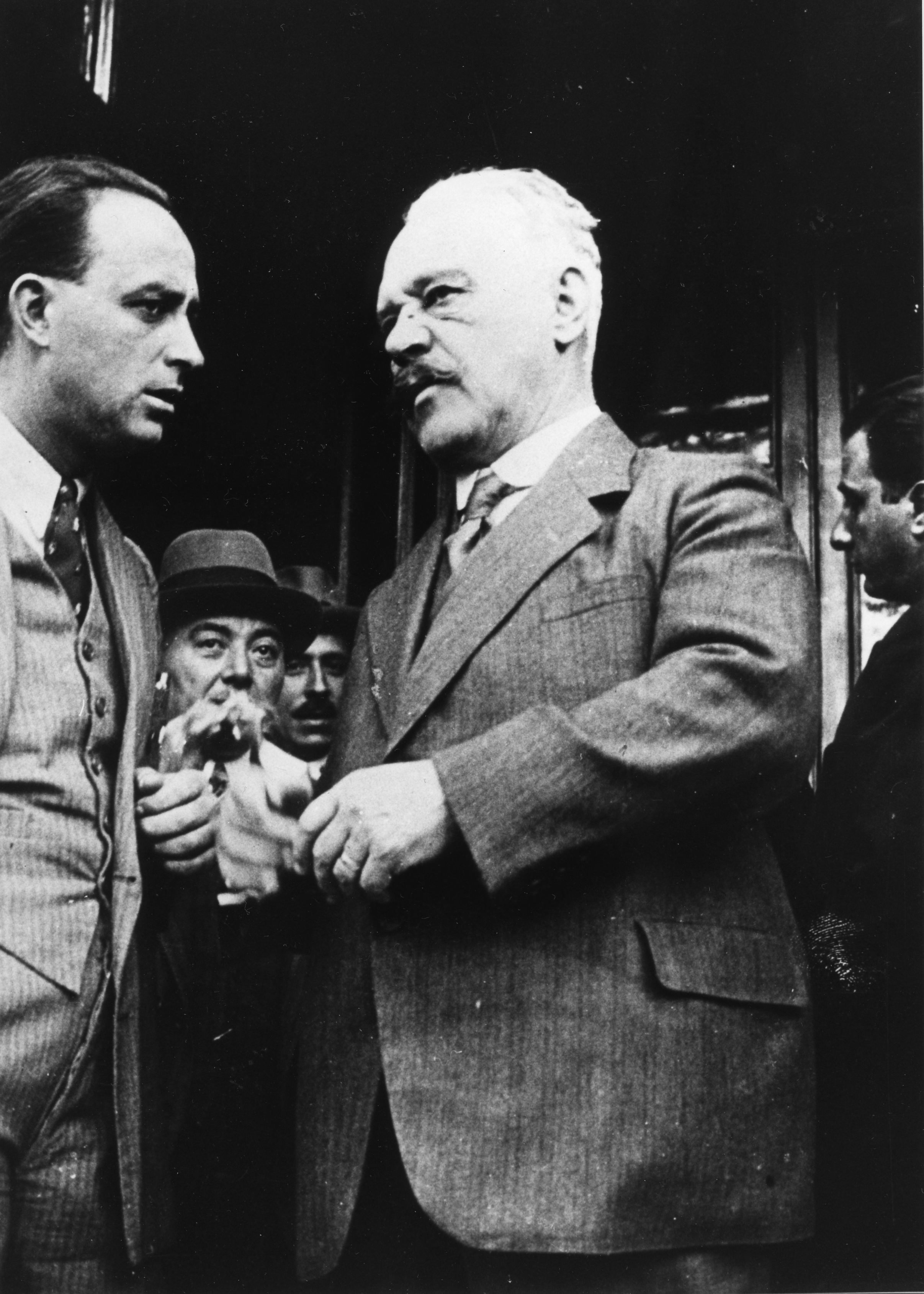}
\caption{ Fermi and Sommerfeld, Rome 1931 (AIP Emilio Segrè Visual Archives, Segrè Collection).}\label{fig8}
\end{figure}

As Rasetti recalls, the consequences of this conference were enormous in that they demonstrated to Italian academics how important Fermi was. {\it``Fermi was very famous in Germany and Copenhagen at the time of the conference, but in Italy his greatness was appreciated by only a handful of people.''} \citep{kuhn}.   Indeed Einstein had written to Lorentz in June of 1926, suggesting  Fermi and
Langevin as speakers on quantum statistics, at the fifth Solvay
Conference. Fermi did not go to Brussels and the talk on statistics was given by Langevin.

Another  physical application of the statistics, that  to deal with the internal electrons of a heavy atom, was performed  by Fermi himself and independently by Thomas \citep{thomas1927,fermi1927}.

Fermi-Dirac statistics have important applications not only in all areas of physics and astrophysics but also in technology, as they explain, for example, the
behaviour of semiconductors, which are the basis of transistors and
all modern electronics. 

The relevance
of Fermi-Dirac statistics in semiconductors was emphasised by John Bardeen,
winner, with William Shockley and Walter Brattain, of the Nobel Prize for his research on semiconductors and the invention of transistors, in his Nobel lecture in
1956\footnote{Bardeen 1956 Nobel lecture, https://www.nobelprize.org/uploads/2018/06/bardeen-lecture.pdf}:
{\it``Occupancy of the levels is given by the position of the Fermi level, $E_F$.
The probability, $f$, that a level of energy $E$ is occupied by an electron is given
by the Fermi-Dirac function:
\be
f=\f 1 {1+e^{{ {(E-E_F)}/{\dd kT}}}}
\ee}

It is interesting to note that Bardeen  was the only researcher to win two Nobel Prizes
in Physics, and that his second Nobel Prize,  received in 1972 with Leon Cooper and Robert Schrieffer for his
theory of superconductivity in the metals at low temperature,  was also based on Fermi-Dirac statistics \citep{Bardeen:1957kj,Bardeen:1957mv}. The key ingredient, for explaining this phenomenon, was based on the attractive interaction between pair of electrons belonging to the Fermi surface, mediated by phonon exchange. The superconductivity  was also suggested to happen in  nuclear matter and neutron stars by Bohr \citep{Bohr:1958zz};
%and Migdal \citep{Migdal:1959noc}
in this case the pairing is between  the  nucleons (protons and neutrons). At extremely high densities, as in neutron stars, also the quarks are expected to form pairs, as electrons  in a superconductor at low temperature. This high-density quark behaviour is called {\it color superconductivity} \citep{Alford:1997zt}.

\section*{Acknowledgements}
 The authors would like to thank Luisa Bonolis for a critical reading and stimulating suggestions, in particular the Heisenberg oral history interview by Kuhn.
%Imports the bibliography file "sample.bib"
\bibliography{sample1}

\end{document}